\documentclass[pra,twocolumn,showpacs,superscriptaddress,floatfix]{revtex4}

\let\a=\alpha \let\b=\beta  \let\g=\gamma  \let\d=\delta \let\e=\varepsilon
\let\z=\zeta  \let\h=\eta   \let\th=\theta  \let\l=\lambda
\let\m=\mu    \let\n=\nu    \let\x=\xi         \let\r=\rho
\let\s=\sigma \let\t=\tau   \let\f=\varphi 
\let\ch=\chi  \let\ps=\psi   
\let\G=\Gamma \let\D=\Delta  \let\Th=\Theta\let\L=\Lambda 
    \let\Si=\Sigma     
\let\O=\Omega 

\font\tenmib=cmmib10\font\sevenmib=cmmib7\font\fivemib=cmmib5%
\mathchardef\Ba   = "050B  
\mathchardef\Bb   = "050C  
\mathchardef\Bg   = "050D  
\mathchardef\Bd   = "050E  
\mathchardef\Be   = "0522  
\mathchardef\Bee  = "050F  
\mathchardef\Bz   = "0510  
\mathchardef\Bh   = "0511  
\mathchardef\Bthh = "0512  
\mathchardef\Bth  = "0523  
\mathchardef\Bi   = "0513  
\mathchardef\Bk   = "0514  
\mathchardef\Bl   = "0515  
\mathchardef\Bm   = "0516  
\mathchardef\Bn   = "0517  
\mathchardef\Bx   = "0518  
\mathchardef\Bom  = "0530  
\mathchardef\Bp   = "0519  
\mathchardef\Br   = "0525  
\mathchardef\Bro  = "051A  
\mathchardef\Bs   = "051B  
\mathchardef\Bsi  = "0526  
\mathchardef\Bt   = "051C  
\mathchardef\Bu   = "051D  
\mathchardef\Bf   = "0527  
\mathchardef\Bff  = "051E  
\mathchardef\Bch  = "051F  
\mathchardef\Bps  = "0520  
\mathchardef\Bo   = "0521  
\mathchardef\Bome = "0524  
\mathchardef\BG   = "0500  
\mathchardef\BD   = "0501  
\mathchardef\BTh  = "0502  
\mathchardef\BL   = "0503  
\mathchardef\BX   = "0504  
\mathchardef\BP   = "0505  
\mathchardef\BS   = "0506  
\mathchardef\BU   = "0507  
\mathchardef\BF   = "0508  
\mathchardef\BPs  = "0509  
\mathchardef\BO   = "050A  
\mathchardef\BDpr = "0540  
\mathchardef\Bstl = "053F  

\newdimen\xshift \newdimen\xwidth \newdimen\yshift \newdimen\ywidth

\def\ins#1#2#3{\vbox to0pt{\kern-#2pt\hbox{\kern#1pt #3}\vss}\nointerlineskip}

\def\eqfig#1#2#3#4#5{
\par\xwidth=#1pt \xshift=\hsize \advance\xshift
by-\xwidth \divide\xshift by 2
\yshift=#2pt \divide\yshift by 2
{\hglue\xshift \vbox to #2pt{\vfil
#3 \includegraphics{#4.eps}
}\hfill\raise\yshift\hbox{#5}}}

\def\V#1{{\bf #1}}
\def\lis#1{{\overline#1}}

\font\titolo=cmbx12
\font\msytw=msbm10%
\def\RRR{\hbox{\msytw R}}
\def\tende#1{\,\vtop{\ialign{##\crcr\rightarrowfill\crcr
 \noalign{\kern0pt\nointerlineskip} \hskip3.pt${\scriptstyle
 #1}$\hskip3.pt\crcr}}\,}
\def\EE{{\cal E}}
\def\HH{{\cal H}}\def\NN{{\cal N}}
\def\defi{\,{\buildrel def\over=}\,}

\let\0=\noindent\def\*{\vskip2mm}

\def\XX{{\cal X}}
\def\EE{{\cal E}}
\def\HH{{\cal H}}\def\NN{{\cal N}}\def\CC{{\cal C}}

\font\msytw=msbm10
\font\msytww=msbm8 
\def\RRR{\hbox{\msytw R}}\def\rrr{\hbox{\msytww R}}

\def\media#1{{\langle \,#1\,\rangle}}

\def\be{\begin{equation}}\def\ee{\end{equation}}

\let\0=\noindent
\def\Eq#1{{\label{#1}}}%
\def\equ#1{(\ref{#1})}

\def\iniz{\setcounter{equation}{0}}

\begin{document}

\centerline{\titolo
Frictionless thermostats and intensive}
\centerline{\titolo  constants of motion}
{\vskip3mm}

\centerline{G. Gallavotti${}^*$ and E. Presutti${}^@$}
\centerline{${}^*$ Fisica-INFN Roma1, ${}^@$ Matematica Roma2}
\centerline{\today}

{\vskip3mm}
\noindent {\bf Abstract}: {\it Thermostats models in space dimension
 $d=1,2,3$ for nonequilibrium statistical mechanics are considered and
 it is shown that, in the thermodynamic limit, the evolutions admit
 infinitely many constants of motion: namely the intensive
 observables.}  {\vskip3mm}

\setcounter{equation}{0}
\setcounter{section}{0}
\def\SEC{Thermostats}
\section{Thermostats}\label{sec1}
\iniz

 Systems in nonequilibrium statistical mechanics have to be thermostatted,
 and the most realistic thermostats are infinite (i.e., very large) systems
 initially in equilibrium.  The present paper discusses a small (classical)
 system interacting with fairly realistic infinite (classical) thermostats
 of dimension $d=1,2$, or $3$.  This can be attacked via recent
 difficult results on the time evolution of infinite systems in dimension
 $d=1,2$, or $3$.  It is assumed that the initial equilibrium states of the
 thermostats are away from phase transitions.  Some technical assumptions
 on the interactions are also made.  The result obtained here may
 then be expressed physically as follows: at any finite time each
 thermostat remains close to equilibrium in the sense that its global
 temperature remains the same, and this is also true for other intensive
 thermodynamic variables.  If an infinite time limit were taken the
 situation would probably be quite different (and nontrivial only in
 dimension $d=3$) but this is a hard problem, and not tackled in the
 present paper.

The class of models that we shall investigate is when particles of a
 {\it test system}, in a container $\O_0$, and $\n$ other particles
 systems, in containers $\O_1,$ $\ldots,\O_\n$, interact and define a
 model of a system in interaction with $\n$ thermostats, if the
 particles in $\O_1,\ldots,\O_\n$ can be considered at fixed
 temperatures $T_1,\ldots,T_\n$.

A representation of the system is in Fig.1:

\eqfig{200}{86}{%
\ins{40}{15}{%
$x=(\V X_0,\dot{\V X}_0,%
\V X_1,\dot{\V X}_1,\ldots,\V X_\n,\dot{\V X}_\n)$}
}{fig1}{}
\vskip-2mm

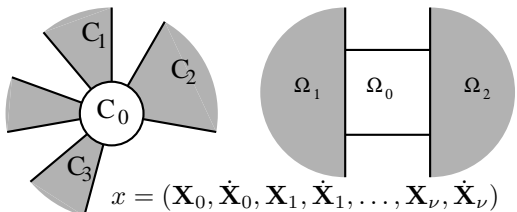
\noindent{Fig.1: \small\it If $d=1,2$ the $1+\n$ finite boxes
  $\O_j\cap\L,\, j=0,\ldots,\n$, are marked $\CC_0,\CC_1,\ldots,\CC_\n$ in
  the first figure and contain $N_0,N_1,\ldots, N_\n$ particles, out of the
  infinitely many particles, with positions and velocities denoted $\V
  X_0,\V X_1,\ldots,\V X_\n$, and $\dot{\V X}_0,\dot{\V X}_1,\ldots,$
  $\dot{\V X}_\n$, respecti\-vely, contained in $\O_j,\,j\ge0$.  The second
  figure illustrates the special geometry that will be considered for
  $d=1,2,3$: here two thermostats, symbolized by the shaded regions,
  $\O_1,\O_2$ occupy half-spaces adjacent to $\O_0$. } \*

\vskip2mm

From the point of view of Physics the temperatures in the thermostats
are fixed.  A natural model, often invoked in the applications,
\cite{FV63}, is to imagine the containers $\O_j,\, j=1,\ldots,\n$, as
infinite and occupied by particles initially in a Gibbs distribution
with given temperatures and densities $T_1,\d_1,\ldots,T_\n,\d_\n$.

To implement the physical requirement that the thermostats have well
defined temperatures and densities the initial data will be imagined
to be randomly chosen with a suitable Gibbs distribution

{\vskip2mm} \noindent {\bf Initial data:} {\it The probability
  distribution $\m_0$ for the random choice of initial data will be,
  if $dx{\,{\buildrel def\over=}\,}\prod_{j=0}^\n\frac{ d\V
  X_j\,d\dot{\V X}_j}{N_j!}$, the limit as $\lis\L\to\infty$ of the
  distributions on the configurations $x\in{\cal H}(\lis\L)$ with $\V
  X_j\in\lis\L$ (see Fig.1),
\be \m_{0,\lis\L}(dx)=const\,\, e^{-H_0(x)}
\,dx\Eq{e1.1} \ee
with $H_0(x)=\sum_{j=0}^\n \b_j\, (K_j(\dot{\V X}_j)-\l_j N_j+ U_j(\V
X_j))$ and $ \b_j{\,{\buildrel def\over=}\,} \frac1{k_BT_j}>0,\,\l_j\in
\RRR, j>0$; the values $ \b_0=\frac1{k_BT_0}>0,\l_0\in\RRR$ will also
be fixed.}
\*

The values $\b_0,\l_0$ bear no particular physical meaning because the test
system is kept finite.  Here $\Bl=(\l_0,$ $\l_1,\ldots\l_\n)$ and ${\bf
  T}=(T_0,T_1,\ldots, T_\n)$ are fixed {\it chemical potentials} and {\it
  temperatures}, and $\lis\L$ is a ball centered at the origin and of
radius $r_0$. The $K_j(\dot{\V X}_j),U_j(\V X_j)$ are kinetic and potential
energies of the particles in $\O_j$ (see below for the conditions on the
potentials).

The distribution $\m_0$ is interpreted as a Gibbs distribution $\m_0$
obtained by taking the ``termodynamic limit'' $\lis\L\to\infty$.
At time $0$ we switch on the interaction between the particles in
$\Omega_0$ and those in the thermostats $\Omega_j$, $j>0$.  The measure
$\mu_0$ is not time invariant under the corresponding dynamics (existence
of dynamics in infinite systems is not at all a trivial issue, as already
revealed by the theory of the evolution in the infinite space
\cite{MPP975} and as it will be discussed later) and we need to:

\0(i) define the temperatures of the thermostats
(which are outside equilibrium);

\0(ii) prove that the ``macroscopic'' property of the thermostats of having
given densities and temperatures remains when the system evolves in time.

If  $p_j(\b,\l;\lis\L) {\buildrel def\over=} \frac1{\b\,|\O_j\cap\lis\L|}\log
Z_j(\b,\l)$ with

\be Z_j(\b,\l)=\sum_{N=0}^\infty \int \frac{dx_N}{N!} e^{-\b(-\l
  N+K_j(x_N)+U_j(x_N))}
\Eq{e1.2}\ee
where the integration is over positions and momenta of the $N$
particles in $\lis\L\cap\O_j$ then we shall say that (at least at time 0) 
the thermostats
have pressures $p_j(\b_j,\l_j)$, densities $\d_j$, temperatures $T_j$,
energy densities $e_j$, and potential energy densities $u_j$, for
$j>0$, given by equilibrium themodynamics, {\it i.e.}:

\begin{eqnarray}
p_j(\b,\l)\defi\lim_{\lis\L\to\infty} p_j(\b_j,\l_j,\lis\L)\kern3cm&&\nonumber\\
\d_j=-\frac{\partial
p_j( \b_j,\l_j)}{\partial\l_j},k_BT_j= \b_j^{-1}\kern2.7cm&&\label{e1.3}\\
e_j=-\frac{\partial \b_jp_j(\b_j,\l_j)}{\partial \b_j}-\l_j\d_j,
\qquad u_j=e_j-\frac{d}2 \d_j\b_j^{-1}\hfill\nonumber&&
\end{eqnarray}
which are the relations linking density $\d_j$, temperature
$T_j=(k_B\b_j)^{-1}$, energy density $e_j$ and potential energy
density $u_j$ in a grand canonical ensemble and in absence of phase
transitions in correspondence of the parameters $(\b_j,\l_j)$, for
$j>0$.
\*

\0{\it Remark:} (1) notice that the limit defining $p_j$ does not
depend on the shape of $\O_j$ and coincides with the usual definition
of pressure in the thermodynamic limit in the sense of Van Hove,
\cite{Ru969}.
\\
(2) As usual in Physics we could define density, energy density and temperatures
in single configurations $x$ as

\be \lim_{n\to\infty}\Big( \frac{N_{j,\L_n}(x)}{|\L_n\cap \O_j|},\
\frac{U_{j,\L_n}(x)}{|\L_n\cap \O_j|},\
 \frac{K_{j,\L_n}(x)}{N_{j,\L_n}(x)}\Big)\Eq{e1.4}\ee
provided the limit exists.
\\
(3) By the Birkhoff theorem applied to systems in the full space ${\bf
R}^d$, the limits exist with probability $1$ for any translational invariant
infinite-volume Gibbs measure ({\it i.e.} a DLR distribution,
\cite{LR969}).  Moreover under an additional assumption of ``extremality''
the limits are almost surely the same for all $x$.  By suitable assumptions
on the parameters $\beta_j$ and $\lambda_j$, stated later in this section,
we shall see that the limits in Eq.\equ{e1.4} exist with $\mu_0$
probability $1$ and are  equal to the values
in Eq.\ref{e1.3}.
\*

Time independence of the intensive observables (in particular those in
Eq.\equ{e1.4}) is the central issue in this paper.  Even if the evolution is
defined with only $H_0$, {\it i.e.} no interaction between $\Omega_0$ and the
thermostats so that $\mu_0$ is time-invariant, yet, in general, one can
only conclude that along ``typical trajectories'' the intensive observables
are constant at countably many times (for instance at all rational times).

However under our assumptions on $\beta_j$ and $\lambda_j$ (essentially
absence of phase transitions) and in the interesting case when the
interaction between $\Omega_0$ and the thermostats is switched on then, by
choosing the initial configurations with $\mu_0$ probability $1$, we shall
prove that the intensive observables keep the same initial value at
all finite times.  This justifies our terminology to call thermostat the
systems $\Omega_j$, $j>0$.
\*

\0{\bf Hypotheses:} {\it In the geometries of Fig.1 suppose:
\\ (1) $\m_0$ satisfies the DLR equations and that \\ (2) the
thermostats pressures $p_j(\b,\l)$ are differentiable in $\b,\l$ at
$\b_j,\l_j$, $j=1,\ldots,\n$.} \*

It is essential that the ``macroscopic'' property of the
thermostats, of having given densities and temperatures, remains when
the system evolves in time.

Evolution is defined via equations of motion: since we are dealing
with infinitely many particles it will be defined by first considering
the motion of the particles initially contained in some ball $\L$
keeping the particles outside $\L$ fixed. Such motion $x\to
S_t^{(\L)}x$ is called $\L$--{\it regularized}: then we shall consider
the limit as $\L\to\infty$.

The regularization boxes $\L$ will be (for simplicity) {\it balls $\L_n$
centered at the origin $O$ and with radius $2^n r_\f$}, with $r_\f$
equal to the range of the interparticle potential, and particles will
be reflected at the boundary of $\L_n$. The limit motion reached as
$n\to\infty$ will define the thermodynamic limit motion.

The $\L_n$--regularized equations of motion will be

\begin{eqnarray}
m\ddot{\V X}_{0i}=-\partial_i U_0(\V X_0)-\sum_{j>0}
\partial_i U_{0,j}(\V X_0,\V X_j)+\BF_i(\V X_0)\nonumber\\
m\ddot{\V X}_{ji}=-\partial_i U_j(\V X_j)-
\partial_i U_{0,j}(\V X_0,\V X_j)\label{e1.5}\end{eqnarray}
(see Fig.1) where:
\*\0(1) the first label, $j=0$ or $j=1,\ldots,\n$, refers
(respectively) to the test system or to a thermostat, while the second
indicates the derivatives with respect to the coordinates of the
points located in the corresponding container {\it and in the
regularization box $\L_n$} (hence the labels $i$ in the subscripts
$(j,i)$ have $N_j\,d $ values).
\\(2) The forces $\BF(\V X_0)$ are,
positional, {\it nonconservative}, smooth stirring forces, possibly
vanishing; the other forces are conservative and generated by a pair
potential $\f$, with range $r_\f$, which couples all pairs in the same
containers and all pairs of particles one of which is located in
$\O_0$ and the other in $\O_j$ ({\it i.e.} there is {\it no direct
interaction} between the different thermostats).
\\(3) Furthermore particles are repelled by the boundaries of the
containers by a conservative force of potential energy $\ps$,
diverging with the distance $r$ to the walls as $r^{-\a}$, for some
$\a>0$, and of range $r_\ps\ll r_\f$. The potential energies will be
$U_j(\V X_j), \,j\ge0$, and $U_{0,j}(\V X_0,\V X_j)$, respectively
denoting the internal energies of the various systems and the
potential energy of interaction between the test system and the
thermostats:

\begin{eqnarray}
U_j(\V X)=\sum_{q\in\V X_j}\ps(q)+\sum_{(q,q')\in\V
  X_j,q\in\L}\f(q-q')\nonumber
&&\\
U_{0,j}(\V X_0,\V X_j)=\sum_{q\in\V X_0,q'\in\V X_j}\f(q-q')\kern1cm&&
\label{e1.6}\end{eqnarray}
The potentials $\f,\ps$ have been chosen $j$--independent for
simplicity.
\\
(4) The equations are formally defined also in the {\it phase
space} $\HH$ of the locally finite configurations $x=(\ldots, q_i,\dot
q_i,\ldots)_{i=1}^\infty$
\be x=(\V X_0,\dot{\V X}_0, \V X_1,\dot{\V X}_1, \ldots, \V
X_n,\dot{\V X}_n)=(\V X,\dot{\V X})
\Eq{e1.7}\ee
with $\V X_j\subset\O_j$, hence $\V X\subset\O=\cup_{j=0}^n \O_j$, and
$\dot q_i\in \RRR^d$; in every ball $\Si(r')$ of radius $r'$ and
center at the origin $O$, fall a finite number of points of $\V X$.
\*

Infinite systems are idealizations not uncommon in statistical
mechanics. But we take it for granted that they must be considered as
limiting cases of large yet finite systems. This leads to several
difficulties: one is immediately manifest if one remarks that the
equations of motion Eq.\ref{e1.5} do not even admit an obvious
solution in $\HH$.

Dynamics is well defined with $\m_0$-probability $1$ because if
$d=1,2,3$ the $\L_n$--regularized equations with data $x$ admit, with
$\m_0$--probability $1$, a limit $S_tx\defi $ $ \lim_{\L_n\to\infty}
S^{(\L_n)}_tx$ for all $t>0$: a precise statement is in theorem 4
below (proved in \cite[theorems 6,7]{GP009}, for $d=1,2$, and
in \cite[Theorem 1]{GP009a} for $d=1,2,3$).

\* Since the Eq.\ref{e1.5} are Newton's equations we shall call the model a
   {\it frictionless} thermostats model: this is to contrast it with other
   thermostats models in which artificial ``frictional'' forces are
   introduced to make it possible for the system to reach a stationary
   state. In models with friction {\it entropy production} ({\it generated
     in the thermostats by their interaction with the system}) due to the
   evolution is naturally defined in terms of the phase space contraction:
   it is therefore interesting to see that even in absence of friction
   entropy production occurs and actually it can be identified, in the
   thermodynamic limit, with the same quantity that would arise in
   thermostats realized via artificial frictional forces. The latter are
   widely studied in the numerical simulations as approximations to
   infinite systems in a thermodynamic limit, because it is not possible to
   simulate really infinite systems. See Sec.\ref{sec5}\*

An important question is whether time evolution changes the
configuration $x$ into $S_t x$ {\it but keeps the temperatures and
densities of the thermostats constant at least with
$\m_0$--probability $1$ and for any finite time}. This is part of the
more general question whether the spatial average of an intensive
observable remains constant in time.

A simple, partial but quantitative, formulation is in
terms of the number $N_{j,\L}(S_tx)$ of particles of $S_tx$, of the
kinetic energy $K_{j,\L}(S_tx)$ and of the potential energy
$U_{j,\L}(S_tx)$ of the configuration $S_tx$ into which $x$ evolves at
time $t$, inside a ball $\L$ centered at the origin.  Consider, then,
$\forall j>0$ ,the limits (if existent)

\be \lim_{n\to\infty}\Big( \frac{N_{j,\L_n}(S_tx)}{|\L_n\cap \O_j|},\
\frac{U_{j,\L_n}(S_tx)}{|\L_n\cap \O_j|},\
 \frac{K_{j,\L_n}(S_tx)}{|\L_n\cap \O_j|}\Big).\Eq{e1.8}\ee
Under the above ``no phase transition'' assumption on $\m_0$
 we shall
 prove: \*

\0{\bf Theorem 1:} {\it The limits in Eq.\ref{e1.8} exist with
  $\m_0$-pro\-bability $1$ for all times and are time independent.
The limits will be respectively $\d_j,u_j$ and $\frac{d}2 \d_j k_B
T_j$ with $\m_0$--probability $1$, as in Eq.\ref{e1.3}.}
\*

\0{\it Remark:} This shows that the thermostats keep, in the
thermodynamic limit, the same temperature and density that they had in
the initial state: a property that has to be required for the model to
adhere to the physical intuition behind the empirical notion of
thermostats.
Hence density and temperature of the thermostats are {\it constants of
motion}. We shall show that more generally many other intensive
observables are also constants of motion.  \*

\def\SEC{Intensive observables}
\section{Intensive observables}\label{sec2}
\iniz

The definition of an {\it $h_\G$-particles intensive observable} is in
terms of a smooth function $\G(q_1,\dot q_1,\ldots,q_h,\dot q_h)$ on
$R^{2d\,h}$ vanishing for $h\ne h_\G$ and which is ``translation
invariant'', and with ``short range'' $r_\G$.

This means that $\G=0$ if the diameter of $X=(q_1,$ $\ldots,q_h)$
exceeds some $r_\G>0$ and, denoting by $\t_\x (X,\dot X)$ the
configuration $(q_1+\x,\dot q_1,\ldots,q_h+\x,\dot q_h)$, it is $
\G(\t_\x (X,\dot X))=\G(X,\dot X),\,\forall\x\in\RRR^d$.

Given a region $W$ the function $G_W$ of $x=(X,\dot X)$

\be G_W(x)\defi \sum_{Y\subset X\cap W}\G(Y,\dot Y)\Eq{e2.1}\ee
defines a ``{\it local observable}'' in $W\subset R^d$ with potential
$\G$.
\*

We shall say that $G_W$ is an observable {\it of potential type} if
$\G(Y,\dot Y)$ depends only on $Y$, while if it depends only on $\dot
Y$ it will be called {\it of kinetic type}.  \*

Then, if $V_n\defi\O_j\cap\L_n$, $|V_n|\defi{\rm volume}(V_n)$, \*

\0{\bf Definition 1:} {\it The ``local average''of $\G$ on the
configuration $x=(X,\dot X)$ is $|V_n|^{-1}G_{V_n}(x)$. The
corresponding ``intensive observable'' in the $j$-th thermostat is

\be g(x)\defi\lim_{n\to\infty}\frac1{|V_n|}
G_{V_n}(x),\Eq{e2.2}\ee
if the limit exists. Furthermore, given $\m_0$, define the ``intensive
fluctuation'' of $G$ (in the $j$-th thermostat)

\be
\D_{G}(x)\defi\lim_{n\to\infty}\big(\frac1{|V_n|}
G_{V_n}(x)-\m_0(\frac1{|V_n|} G_{V_n})\big)\label{e2.3}\ee
\vskip-5mm
$$\kern-27mm\defi\lim_{n\to\infty} \D_{G,V_n}(x),$$
if the limit exists. }
\*

\0{\it Remark:} The notation requires keeping in mind that $G_{V_n}$
depends also on $j$ (because $V_n=\O_j\cap\L_n$): however for
simplicity of notation the labels $j$ on $V_n$ and $G_{V_n}$ will not be
marked.  \*

Properties of intensive observables can be derived from various
assumptions on the {\it initial} distributions of the particles in the
various regions $\O_j$ which, we recall, are distributed independently
over $j=1,\ldots,\n$ and depend on the $\n$ pairs of parameters
$\b_j,\l_j$.

The simplest assumption is perhaps the uniqueness of the tangent plane
to the graph of the pressure in various directions, which could for
instance be insured by the uniqueness of the translation invariant
states of our particles system with parameters $\b_j,\l_j$.

Let $G$ be an observable of potential or kinetic type; and suppose
that $H_{0,\L, \G}(x)$ $ \defi H_{0,\L}(x)+\th G_{\L}(x)$ is
superstable for $|\th|$ small enough ({\it i.e.} there exist constants
$a>0,b\ge0$ such that for all balls $\L$ it is $H_{0,\L, \G}(x)\ge a
N^2/|\L|-bN$ for all configurations $x=(X,\dot X)$ with $N$ particles
and with $X\subset \L$ and $\forall\, |\th|\le\th_0$ for some
$\th_0>0$.  We call $G$ an ``{\it allowed observable}''. For such
observables it is possible to define, for $|\th|$ small, the
``pressure''

\be P(\th)=\lim_{\L\to\infty}\frac1{|V|}\log
\frac{Z_j(\th)}{Z_j(0)}\Eq{e2.4}\ee
with $Z_j(\th)$ given by Eq.\ref{e1.2} with the energy $\th\,
G_{V}(x)$ added in the exponential. It is $P(0)\equiv0$.

It is important to stress that $ P(\th)$ is, in the geometries in
Fig.1 considered here, {\it independent} of the special geometry
considered for the $\O_j$ as long as the conical containers have
$d$--dimensional shape ({\it i.e. they contain balls of arbitrarily
large radius}).

In this context we can derive the following result: \*

\0{\bf Theorem 2:} { \it Let $G$ be an allowed observable of potential
  or kinetic type. If $P(\th)$ is differentiable at $\th=0$, then with
  $\m_0$--probability $1$ the limit as $|V_n|\to\infty$ of
  $\frac1{|V_n|}G_{V_n}(S_tx)$ exists $\m_0$--almost everywhere and is
  $t$-independent.}

\* \0{\it Remarks:} (1) The differentiability assumption of $P(\th)$
has the meaning of uniqueness of the tangent plane to the graph of the
pressure $p$ ``in the direction of $G$'': such uniqueness is a
``generic'' property, see \cite{GM967} for the lattice gas case.
\vskip2mm\0(2) The superstability of $H_{0,\lis\L}(x)+\th G_{\lis\L}(x)$
is a very strong condition: it is certainly satisfied if
\vskip2mm\0
\hglue2mm (i) $\G(X,\dot X)=1$ for
$|X|=1$ and $0$ otherwise, or if
\vskip2mm\0
\hglue2mm (ii) $\G(X,\dot X)=\frac12\dot q^2$ for $|X|=1$
and $0$ otherwise, or  if
\vskip2mm\0
\hglue2mm (iii) $\G(X,\dot X)=0$ unless $X=(q,q')$ and in
such case \hglue8mm $\G(q,q')=\f(q-q')$,
\\
therefore
theorem 1 is a corollary of theorem 2.
\*

We also expect that the intensive observables will have very small
probability of being appreciably different from their average values,
and precisely a probability bounded above by an exponential of the
volume $|\L_n|$. This will mean that the observable $G$ satisfies a
kind of {\it large deviations property}: \*

\0{\bf Theorem 3}: {\it Under the assumptions of theorem 2 the
$\m_0$--probability that the fluctuation $\D_{G,\L_n}(S_tx)$ differs
from $0$ by more than $\e>0$ tends to $0$ exponentially fast in
$|V_n|$ as $n\to\infty$, $\forall\,\e\,>0$.}  \* \*

\0{\it Remark:} The assumptions in theorems 2,3 are satisfied by many
observables in the Mayer expansion convergence region in the plane
$\l_j,\b_j$, \cite{GBG004}. They are also believed to be satisfied
quite generally for observables generated by a potential $\G$. In
particular they hold generically if $\G$ is a linear combination of
the potentials (i),(ii),(iii) in remark (2) above.

The proof of theorems 2,3 are presented in Sec.IV.

\def\SEC{Time evolution}
\section{Time evolution}\label{sec3}
\iniz

A quantitative existence theorem of the dynamics can be conveniently
formulated in terms of the quantities $v_1\defi
\sqrt{2\f(0)/m},\,r_\f$ and $W,\NN,v_1,\|x_1\|$ defined as

\begin{eqnarray}
W(x;\x,R) \defi\frac1{\f(0)}\sum_{q_i\in {\cal
B}(\x,R)}\big(\frac{m\dot q_i^2}2\kern2cm\nonumber\\
+\frac12\sum_{j; j\ne
i}\f(q_i-q_j)+\ps(q_i)+\f(0) \big),\kern1cm\nonumber\\
\NN_\x(x)\defi \hbox{number particles within $r_\f$
of $\x\in\RRR^d$},\\ ||x_i-x'_i||\defi |\dot
q_i-\dot q'_i|/v_1+|q_i-q'_i|/r_\f\label{e3.1}\nonumber\end{eqnarray}
Let $\log_+ z\defi\max\{1,\log_2|z|\}$, $g_\z(z)=(\log_+ z)^\z$ and

\be \EE_\z(x)\defi \sup_{\x}
\sup_{R> g_\z(\x/r_\f)} \frac{W(x;\x,R) }{R^d}.
\Eq{e3.2}\ee
Call ${\cal H}_\z$ the configurations in $\cal H$ with

\be
(1)\ \ {\cal  E}_\z(x)<\infty\kern3cm
\label{e3.3}\ee
$$(2)\ \ \frac{N(j,\L_n)}{|\L_n\cap\O_j|},\
\frac{U(j,\L_n)}{|\L_n\cap\O_j|},
\ \frac{K(j,\L_n)}{|\L_n\cap\O_j|}
\tende{n\to\infty}\,
\d_j,u_j,\frac{d\,\d_j}{2\b_j}$$
with $\L_n$ the ball centered at the origin and of radius $2^n\, r_\f$
$\d_j,u_j,T_j,$ given by Eq.({e1.8}) if $N(j,\L_n)$, $U(j,\L_n),K(j,\L_n)$
denote the number of particles and their internal potential or kinetic
energy in $\O_j\cap\L_n$. Each set ${\cal H}_\z$ has $\m_0$-probability
$1$ for $\z\ge1/d$, \cite{Ru970,MPP975,FD977,CMP000}. Then:
\*

\0{\bf Theorem 4:} {\it Let $d\le3$, then $\HH_{1/d}$ has
  $\m_0$--probability $1$ and $S_t x$
  exists for $\m_0$--almost all $x\in\HH_{1/d}$ and $\forall
  t\ge0$. Given (arbitrarily) a time $\Th>0$,
 if $\EE\defi\EE_{1/d}(x)$, and $|q_i(0)|\le
  2^k r_\f$ there are $c=c(\EE,\Th)<\infty, c'=c'(\EE,\Th)>0$ such that
$\forall\, n\ge k$ and $\forall\, t\le\Th$

\begin{eqnarray}
|\dot q_i(t)| \le c\, v_1 \, k^{\frac12} ,\nonumber\kern5cm\\
{\rm distance}( q_i(t), \partial(\cup_j \O_j))\ge\, c'\, r_\f\,
k^{-\frac1\a}\nonumber\kern2cm\\
\NN_\x(S_tx)\le\,  c\, k^{1/2}\kern5cm\\
\|(S_tx)_i-(S^{(n)}_tx)_i\|\, \le\,  e^{-c'2^{n/2}} ,\ n>k.
\label{e3.4}\nonumber\kern1.7cm\end{eqnarray}
}

\* This is proved in \cite[theorem 7]{GP009} for $d=2$ and in
\cite{GP009a} for $d=3$ (the latter reference covers also the case
$d=2$ via a somewhat different approach).

Remark that the theorem {\it does not state} that the second of
Eq.\ref{e3.3} holds: in \cite{GP009,GP009a} it is however proved, in
addition to theorem 4, the weaker statement that the
$\liminf$ of $\frac{K_{j,\L_n}(S_tx)}{|\O_j\cap\L_n|}$ is not smaller
than $ \frac12$ of the corresponding {\it r.h.s.}; and the same is
true for the other two quantities in Eq.\ref{e3.3}.

A corollary of the main results of this paper will be that the limit
relations in Eq.\ref{e3.3} will hold for all $t>0$.

\def\SEC{Constants of motion}
\section{Constants of motion}\label{sec4}
\iniz

Let $\G$ be an $h$-points local observable of potential type,
$V_n=\O_j\cap\L_n$. Under the assumptions of theorem 2 we first show
that $\lim_{n\to\infty} |V_n|^{-1}\media{G_{\L_n}}_{\m_0}=g$ exists.

Define $P_n(\th)\defi $ $\frac1{|V_n|}\log \media{e^{-\th
    G_{V_n}}}_{\m_0}$: this is smooth and convex in $\th$ and its
    unique derivative at $\th=0$ is $g_n\defi
    \frac1{|V_n|}\m_0(G_{V_n})$; therefore, remarking that $P(0)=0$, it
    satisfies $P_n(\th)\,\ge \,\th\, g_n$.

The limit $P(\th)$ as $n\to\infty$ of $P_n(\th)$ is the same that
would be obtained if $V_n$ was replaced by the full ball $\L_n$ and
filled with particles at temperature $\b_j^{-1}$ and chemical
potential $\l_j$.

Any convergent subsequence $g_{n_i}$ defines therefore a coefficient
$g$ with the property $P(\th)\ge \th g$. Hence, by the assumed
uniqueness of the tangent to $P(\th)$ at $\th=0$, it follows that $g$
is uniquely determined thus implying that the limit $g\defi
\lim_{n\to\infty}g_n$ exists.

Let $g_n=\media{|V_n|^{-1}G_{\L_n}}_{\m_0}$ and, given $\g>0$, let
$\XX_{E,\g,n}$ to be the set of points in $\HH_{1/d}$ with $\EE(x)\le
E, \,G_{\L_n}(x)<(g_n+\frac12\g)|V_n|$ and which, under the evolution,
reach in a time $\t_{\g,n}(x)\le\Th$ and for the first time, a point
of the surface

\be \Si_{n,\g}\defi \{\,x\,|\,|V_n|^{-1}G_{\L_n}(x)=
(g_n+\g)\,\}.\Eq{e4.1}\ee

If for all $E$ and for all small $\g>0$ it is $\sum_n
\m_0(\XX_{E,\g,n})<+\infty$ then it will be $\limsup_{n\to\infty}
|V_n|^{-1}G_{\L_n}(S_tx)\le g$, with $\m_0$--probability $1$ (by
Borel--Cantelli's estimate); changing $\G$ into $-\G$ it will follow,
again with $\m_0$--probability $1$, that the $\liminf$ is $\ge g$:
notice that the change in sign of $\G$ is possible by the conditon on
$G$ to be an ``allowed observable'' , as introduced before
Eq.\ref{e2.4}.

This remains true if for all small $\g$ there exists $\g_n\in [\g,2\g]$
such that $\sum_n $$\m_0(\XX_{E,\g_n,n})<+\infty$.

If $x\in \XX_{E,\g,n}$ the phase space contraction, when phase space
volume is measured by $\m_0$, within time $t$ is, \cite{GP009,GP009a},

\be s(x,t)=\int_0^{t}\,\big(\sum_{j\ge0}
\b_j Q_j(\t)+ \b_0 L_0(\t)\big)\,d\t\Eq{e4.2}\ee
where $Q_j(t)\defi \dot{\V X}_j(t)\cdot{\bf F}_j,\,L_0(t)\defi\dot{\V
X}_0\cdot\BF(\V X_0(t))$.

By theorem 4, $L_0(t)$ is uniformly bounded as $n\to\infty$, for $0\le
t\le\Th$, by the first of Eq.\ref{e3.2}, by a quantity $C$ (only
depending on $E,n_0,\Th$).

Therefore by a quasi-invariance lemma, \cite{Si972b,MPP975}, \cite[Appendix
  H]{GP009}, the probability $ \m_0(\XX_{E,\g+\e,n})$ can be bounded
$\forall \e\in[\g,2\g]$ by

\be
{C}\int \m_0(dx)\frac{|\widehat G|}{|V_n|}
\,\d(\frac{G_{\L_n}(x)}{|V_n|}-(g_n+\g+\e)) \label{e4.3}\ee
where $\widehat G$ denotes the time derivative (at $t=0$) of
$G_{\L_n}(S_tx)$ (to be computed via the equations of motion)
evaluated on the surface $\Si_{n,\g+\e}$, see Eq.\ref{e4.1}.

Integrating Eq.\ref{e4.3} over $d\e/\g$, $
\m_0(\XX_{E,n,\g_n})$ can be bounded by

\be \frac{{C}}{\g} \int \m_0(dx) \frac{|\widehat G|}{|V_n|}
\,\ch(\g\le \frac{G_{\L_n}(x)}{|V_n|}-g_n\le 2\g), \Eq{e4.4}\ee
with $\widehat G= \sum_{X\subset V_n}$ $ \sum_{q\in X}$ $
\partial_q \G(X)\, \dot q$. By Schwartz' inequality

\be C_2 \g^{-1}\,\m_0(\{x\,:\,\g\le \frac{G_{\L_n}(x)}{|V_n|}-g_n\le
2\g\})^{1/2} \Eq{e4.5}\ee
because from Eq.\ref{e2.1} for $\G$

\be  \m_0(\widehat G^2)^{1/2} \le C_1
|V_n|  \Eq{e4.6}\ee
obtained via superstability bounds, using the Maxwellian distribution
for $\dot q$.

The probability in Eq.\ref{e4.5} is bounded above by Chebischev inequalities
(quadratic or exponential) by both averages

\begin{eqnarray} &I\defi\langle \frac{(G_{\L_n}(x)/|V_n|-g_n)^2}{\g^2}
\rangle_{\m_0},\nonumber\\
&I_\th\defi\langle e^{\th\,
(G_{\L_n}-|V_n|(g_n+\g))} \rangle_{\m_0}\label{e4.7}\end{eqnarray}
$\forall\,\th\ge0$. This implies the existence of $\g_n\in[\g,2\g]$ with:

\be \m_0(\XX_{E,n,\g_n})\le C_3\g^{-1} J(n), \qquad J(n)^2=I,I_\th
\Eq{e4.8}\ee
Therefore we look for assumptions on the thermostats structure ({\it
i.e.} on $\l_j,\b_j,\f$) under which $J(n)$ tends to zero fast enough
making $\sum_n \m_0(\XX_{E,n,\g_n})$ $<\infty$. In this case theorem 2
will follow from Borel-Cantelli's lemma and the arbtrariness of
$\g$.\*

As a consequence of the above bounds, basically following from {\it
the uni\-que\-ness of the tangent plane in the direction $\G$}, the
proof of theorem 2 can be completed as follows.  Fix $\g>0$ and remark
that

\be I_\th=\media{e^{\th U_{\G,V_n}}}_{\m_0} e^{-\th (g_n
 +\g) |V_n|}\le  e^{-\th\g
 |V_n|+\h(\th,V_n)}\Eq{e4.9}
\ee
Continuing the argument leading to the existence of the limit of
$g_n$, at the beginning of the section, the correction term
$\h(\th,V_n)$ is bounded, cas follows: \*
\0(a) $\frac1{|V_n|}\log\media{e^{\th U_{\G,V_n}}}_{\m_0}$ is
$P_n(\th)-P_n(0)$ (notice: $P_n(0)$ $\equiv 0$) and converges to
$P(\th)-P(0)$ as $V_n\to\infty $ for $|\th|\le \th_0$, if $\th_0$ is
small enough so that the potential $\f+\b_j^{-1}\th$ is superstable
$\forall\,|\th|\le\th_0, \,j=1,\ldots,\n$.  By superstability the
limit exists for $|\th|\le\th_0$ and it is a limit of functions
$P_n(\th)$ which are convex for $|\th|\le\th_0$. Hence the limit is
uniform: $|P(\th)-P_n(\th)|\le o(|V_n|)$ for $|\th|\le\th_0$,
\vskip1mm \0(b) the $g_n$ in the exponent in Eq.\ref{e4.7} has just
been shown to be $g_n |V_n|=g |V_n|+o(|V_n|)$, so that $-\th\, g_n$
converges to $-\th\, g$ with an error $\th\, o(|V_n|) $,
\vskip1mm \0(c) $(P(\th)-P(0)-\th\, g) |V_n|$ is (by the uniqueness of
the tangent plane) $o(\th)\,|V_n|$.  Hence

\begin{eqnarray}
&\h(\th,V_n)-\g\th_n |V_n|\le -\frac12\g\th_n
|V_n|\kern2.5cm\label{e4.10}\\
&+\big(-\frac12\g\th_n
+\frac{o(|V_n|)}{|V_n|}+o(\th_n)\big)\,|V_n|\le -\frac12\g\th_n
|V_n|\nonumber\end{eqnarray}
and choosing $\th_n$ tending to $0$ so slowly that the exponent of the
{\it r.h.s.} of \ref{e4.9} tends rapidly to $\infty$, for instance if
$\th_n=\max(\frac1{\log n},\frac1{4\g}\frac{o(|V_n|)}{|V_n|})$, we see
that $I_{\th_n}\tende{n\to\infty}0$ so fast that
$\m_0(\XX_{E,n,\g_n})$ is summable in $n$ implying theorem 2
and of its special case theorem 1.

Theorem 3 also follows from the existence of the limit for $g_n$
because $I_\th$ yields a summable bound on $J$, hence on
$\m_0(\D_{G,\L_n}^2)$.
\*

\0{\it Remarks:} (1)
Uniqueness of the tangent plane can be replaced by assumptions on the
decays of correlations in the distribution $\m_0$ somewhat stronger
than just requiring its extremality among the DLR distributions in the
geometry in Fig.1.
\\
(2) Sufficient estimates can be formulated as follows:
$\r_j(x_1,\ldots,x_n)$ be the $n$--points correlation function in the
$j$--th container: by superstability $\r_j\le C^n$, \cite{Ru970}. If
$x=(q,\dot q)$ and $\x\in\O_j$, extremality of $\m_0$, implies,
\cite{LR969,Ru970}, for $x_1,\ldots,x_n$ and $y_1,\ldots,y_m)$ with
positions in $\O_j$:

\begin{eqnarray}
|\r_j(x_1,\ldots,x_n,\t_\x y_1,\ldots,\t_\x y_m)\nonumber\kern3cm\\
-\r_j(x_1,\ldots,x_n)\r_j(\t_\x y_1,\ldots,\t_\x
y_m)|\tende{\x\to\infty}0\label{e4.11}\end{eqnarray}
{\it Assume that Eq.\ref{e4.11} holds in the stronger sense} that the {\it
l.h.s.} is bounded by $\h_{R,m,n}(\x)$ if the positions of
$x_1,\ldots,x_n$ and $y_1,\ldots,y_m)$ can be enclosed in a ball of
radius $R$.  \*

\0{\bf Theorem 5:} {\it If there is a constant $C_{R,m,n}<\infty$ such that
$\h_{R,m,n}(\x)$ $\le C_{R,m,n}
  |\x|^{-a(R,m,n)}$ with $a(R,m,n)>0$ and if
  $\lim_{\L\to\infty}\frac{1}{|V_n|}\m_0(G_{V_n})=g$ exists, then
  $\lim_{\L\to\infty}$ $\frac1{|V_n|} \D_{G,V_n}(x)=0$ and
  $\lim_{\L\to\infty}\frac{1}{|V_n|} G_{V_n}(S_tx)=g$ with $\m_0$
  probability $1$.} \*

\0{\it Remarks:} (1) Thus if $\m_0$ has a power law
cluster property {\it all} intensive observables admitting an average
value, over space translations, at time $0$ are constants of motion.
\\
(2) With the above assumptions we avoid use of the exponential
Chebishev inequality and we may thus drop the superstability condition
in the definition of the potential $\Gamma$.  We could actually
consider more general observables of the form (in $\Omega_j$)

      \be \lim_{n\to \infty} \frac{1}{|\Omega_j\cap
      \Lambda_n|}\int_{r\in \rrr^d:\tau_r\Delta \subset
      \Omega_j\cap \Lambda_n} \tau_r f (x) dr \Eq{e4.12} \ee
where $f$ is a cylindrical function in $\Delta$ (i.e.\ it does not
depend on the particles outside $\Delta$) and $\tau_r$ denotes
translation by $r$.  If the power law cluster property is satisfied
and $\mu_0$ a.s.\ the limit in \ref{e4.12} exists at time 0, then the
intensive observables \ref{e4.12} are constant of motion under the
assumption that $f$ is smooth and grows at most polynomially with the
number of particles.  \\
(3) The assumption certainly holds in the cluster expansion
convergence region, \cite{Ru969} and \cite[Sec.5.9]{Ga000}, {\it i.e}
high temperature and low density, without extra assumptions.  \*

\0{\it Proof:} Consider the first of Eq.\ref{e4.7} and choose
$\g=\g_n=\frac1n$. The numerator tends to $0$ as $|V_n|^{-a(R,n,n)/d}$
if the potential $\G$ for the observable $G_{V_n}$ vanishes when the
diameter of the set $\{x_1,\ldots,x_n\}$ exceeds $R$.

The estimate Eq.\ref{e4.8} implies that $\frac1{|V_n|}\D_{G,V_n}(S_t
x)$ tends to $0$ with $\m_0$--probability $1$ for all $t\le kt_0$ with
$k$ integer and $t_0>0$ (arbitrarily fixed). Hence if the average of
$\frac{1}{|V_n|}\m_0(G_{V_n})$ exists it exists for all times and has a
time-independent value.

\kern-5mm
\def\SEC{Entropy and thermostats}
\section{Entropy and thermostats}\label{sec5}
\iniz

Entropy production rate (due to the action of the system upon the
thermostats and identified with the rate of {\it their} entropy increase, which
is finite even though the thermostats entropy is infinite because the
thermostats are infinite) is defined in terms of $Q_j=-\dot{\V
  X}_j\cdot\partial_{ \V X_j} U_{0,j}(\V X_0,\V X_j)$, which is the work
per unit time, performed by the test system on the $j$-th thermostat. Since
$Q_j$ is interpreted as the {\it heat} ceded by the system to the
thermostats the entropy production in the configuration $x$ is given by
$\s_0(x)=\sum_{j>0}\b_j Q_j(x)$.  \\ If the volumes in phase space are
measured by the distribution $\m_0$ this quantity differs from the
contraction rate of the phase space volume by $\b_0(\dot Q_0+L_0)\equiv
\b_0(\dot K_0+\dot U_0)$ and $K_0+U_0$ is {\it expected} to stay finite
uniformly in time. If so the statistics of the long time averages of the
phase space contraction rate and of the entropy production rate will
coincide (however this is not proved as the theorems above only concern
what happens in a {\it arbitrarily prefixed but finite} time interval).
\\ 
In other words in the frictionless thermostats model and in the
isoenergetic thermostat models, \cite{GP009a}, the entropy production can
be identified with the phase space contraction, possibly up to a time
derivative of a quantity expected to be uniformly finite in
time. Furthermore the entropy production is the same in both models of
thermostats if the thermodynamic parameters of the thermostats ($
\d_j,T_j,\, j>0$) are the same: this follows from the equivalence theorem
between frictionless and isoenergetic thermostats, \cite[theorem
  1]{GP009a}, which states that under such conditions the microscopic
motions of the two models starting from the same initial condition remain
identical forever with $\m_0$--probability $1$.

Other thermostats can be considered: for instance the isokinetic
thermostats. At a heuristic level analogous conclusions can be
reached, \cite{Ga008d}.

Considering external thermostats as correctly representing the physics of
the interaction of a system in contact with external reservoirs has been
introduced in \cite{WSE004}.  Their analysis was founded on the grounds of
\\
(1) identity, in the thermodynamic limit,
of the evolution with and without thermostats
\\
(2) identity of the phase space contraction of the thermostatted
systems with the physical entropy production (up to a time derivative).

For a more mathematical view see \cite{Ga008d}.

\small
\bibliographystyle{unsrt}

\end{document}